\documentclass[aps,twocolumn,amsmath,amssymb,prl,showpacs]{revtex4-1}
\usepackage{amsfonts, hyperref, times}
\usepackage{graphicx}

\begin{document}
  
\title{Domain-wall dynamics in translationally non-invariant
  nanowires: theory and applications}

\author{O.~A.~Tretiakov}
\author{Y.~Liu}
\author{Ar.~Abanov}

\affiliation{
             Department of Physics \& Astronomy,
	    Texas A\&M University,
            College Station, Texas 77843-4242, USA
}

\date{June 11, 2012}

\begin{abstract}
We generalize domain-wall dynamics to the case of translationally
non-invariant ferromagnetic nanowires. The obtained equations of
motion make the description of the domain-wall propagation more
realistic by accounting for the variations along the wire, such as
disorder or change in the wire shape.  We show that the effective
equations of motion are very general and do not depend on the model
details.  As an example of their use, we consider an hourglass-shaped
nanostrip in detail. A transverse domain wall is trapped in the middle
and has two stable magnetization directions.  We study the switching
between the two directions by short current pulses.  We obtain the
exact time dependence of the current pulses required to switch the
magnetization with the minimal Ohmic losses per
switching. Furthermore, we find how the Ohmic losses per switching
depend on the switching time for the optimal current pulse.  As a
result, we show that as a magnetic memory this nanodevice may be
$10^5$ times more energy efficient than the best modern devices.
\end{abstract}


\maketitle

During last two decades there has been a significant progress in
describing magnetization and, in particular, domain-wall (DW) dynamics
in magnetic nanostructures \cite{Ono99, Yamaguchi04, Beach05,
  Klaui:images05, KlauiPRL05, HayashiPRL2006, Thomas2006, Thomas2007,
  Meier07, Moriya08, Rhensius10, IlgazPRL10, Singh10, KlauiAPL10,
  Krivorotov10, Thomas2010, BeachPRL09, Tatara04, Zhang04, Barnes05,
  Duine07, Tserkovnyak2008, Lucassen09, Duine09, Tretiakov_DMI,
  Tretiakov:losses, Nakatani03, Thiaville05, Martinez07, Martinez09,
  Min10, Brataas2010, Pi2011, YanEPL10, Yang2010}. The interest to
these studies has been inspired not only by the fundamental physics
questions but also by the potential applications for the spintronic
memory and logic nanodevices \cite{Parkin:racetrack08, Allwood01,
  Allwood02}.  However, recently this progress has been staggered due
to the inability to make perfect translationally invariant nanowires
from one side and the difficulty to apply the theories made for
translationally invariant systems to successfully describe some of the
phenomena in the experimental systems from the other side. There have
been attempts to consider analytically and
numerically~\cite{Nakatani03, Thiaville05, Martinez07, Martinez09,
  Min10} the systems with rough surfaces or other disorder but the
general theory for translationally non-invariant magnetic systems is
still lacking.

In this Letter we generalize current and field induced DW dynamics to
the case of translationally non-invariant ferromagnetic
nanowires. This generalization makes the description of the DW motion
more realistic by accounting for variations along the wire, such as
disorder or change in the wire shape. We show that the effective
equations of DW motion are very general and do not depend on the
details of the model Hamiltonian. These equations are the main
theoretical result of this Letter.

\begin{figure}
\includegraphics[width=1\columnwidth]{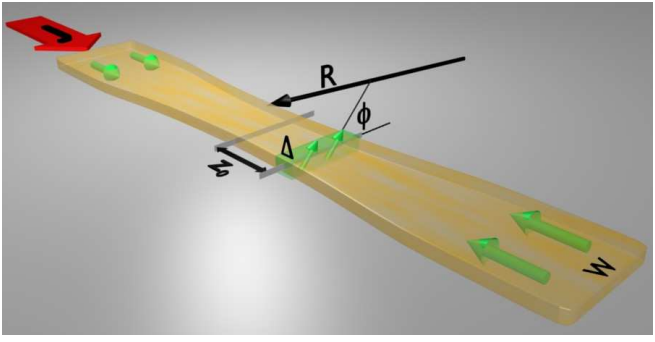}
\caption{(color online) Hourglass-shaped nanostrip as a prototype of a
  magnetic memory nanodevice.}
\label{fig:pulse}
\end{figure}

As an example of the application of this theory, we study
current-induced magnetization switching in a thin hourglass-shaped
nanostrip.  We show that a transverse DW trapped in the middle of a
curved inward nanostrip can serve as a magnetic memory device, see
Fig.~\ref{fig:pulse}. At zero current, the magnetization in the
transverse DW can have either of the two equilibrium directions in the
plane of the nanostrip.  We study the switching between these two
magnetization directions mediated by short current pulses with the
requirement that the DW returns to the initial position after
switching.  The main energy loss during the switching is due to Ohmic
heating of the wire. The minimum energy required per switch depends on
the switching time.  We obtain the exact time dependence of the
current pulses required to switch the magnetization in the most
efficient way (with the minimal Ohmic losses per switching for a given
switching time). Furthermore, we find how the Ohmic losses depend on
the switching time for this optimal current pulse.

The two equilibrium magnetization directions can serve as "zero" and
"one" of a memory bit.  We show that based on this prototype, it is
possible to design a nonvolatile memory device with an extremely short
writing time, which is only limited by the spin-wave frequency.  The
energy required per switch for such a nanodevice is much lower than that
for the state-of-the-art memory devices.

\textit{Equations of motion}. We study the DW dynamics by employing
the Landau-Lifshitz-Gilbert (LLG) equation with current
terms~\cite{Zhang04, Thiaville05}.  In general, a nontrivial solution
of the static LLG equation, $\mathbf{S}_{0}\times \delta
\mathcal{H}_{0}/\delta \mathbf{S}_{0} =0$, has a continuous symmetry.
It means that this solution can be parametrized by an even-dimensional
vector $\boldsymbol{\xi}$ such that
$\mathbf{S}_{0}(z,\boldsymbol{\xi})$ is the solution for any
$\boldsymbol{\xi}$ in a continuous interval.  Then the DW dynamics due
to a correction to Hamiltonian $\mathcal{H}_{0}$ or an electric
current can be described by the time-dependent parameter
$\boldsymbol{\xi}(t)$. The equations for $\boldsymbol{\xi}(t)$ are
called the effective equations of motion. Below we sketch the general
derivation of such effective equations.

To consider the dynamics we assume that there is a perturbation
$\mathbf{h}$, such that the full LLG equation takes the form
\begin{equation}
\label{eq:LLe}
\dot{\mathbf{S}}=\mathbf{S}\times 
\frac{\delta \mathcal{H}_{0}}{\delta \mathbf{S}}+ \mathbf{h},
\end{equation}
where the time is measured in units of the gyromagnetic ratio
$\gamma_0 = g|e|/(2mc)$ and $\mathbf{S}=\mathbf{M}/M$ with $M$ being
the saturation magnetization.  We note that $\mathbf{h}$ is not only a
correction to Hamiltonian and may also contain other contributions
such as dissipation, adiabatic and nonadiabatic current terms.  We
look for a solution of this equation in the form $\mathbf{S}(z,t)
=\mathbf{S}_{0}(z,\boldsymbol{\xi} (t)) +\mathbf{s}$, where the
dependence $\boldsymbol{\xi}(t)$ is weak and $\mathbf{s}$ is small and
orthogonal to $\mathbf{S}_{0}$ at each point.

For current-driven magnetization dynamics the current is assumed
to be uniform in the nanowire.  The correction $\mathbf{h}$ then can
be written as a sum of two terms:
\begin{equation}
\label{eq:h}
\mathbf{h}=\mathbf{S}_{0}\times 
\frac{\delta \mathcal{H}_{\delta }}{\delta \mathbf{S}_{0}}
+\mathbf{h}_{\alpha }, \quad 
\mathbf{h}_{\alpha }=\alpha \mathbf{S}_{0} \times \dot{\mathbf{S}}_{0} 
-j\partial \mathbf{S}_{0}+\beta j\mathbf{S}_{0} \times \partial \mathbf{S}_{0}.
\end{equation}
The first term is the correction to the Hamiltonian,
$\mathcal{H}_{\delta }$, which turns the zero modes into soft modes as
well as couples the magnetization to an external magnetic field. The
second term, $\mathbf{h}_{\alpha }$, is due to dissipation and current
terms. Here $\alpha$ is the Gilbert damping constant, $j$ is an
electric current in the units of velocity, and $\beta$ is the
non-adiabatic spin torque constant.

Taking the scalar product of Eq.~\eqref{eq:LLe} with
$\mathbf{S}_{0}(z)\times \partial_{\xi_{j}}\mathbf{S}_{0}(z)$ and
integrating over the space, one can find the following equation for
the collective coordinates $\xi_i$:
\begin{equation}\label{eq:orthDW}
2\dot{\xi}_{i}= -\epsilon^{ij}\partial_{\xi_{j}}E
-\epsilon^{ij}\int\!\! dz\, \mathbf{S}_{0}(z)\times 
\partial_{\xi_{j}}\mathbf{S}_{0}(z)\cdot \mathbf{h}_{\alpha }(z),
\end{equation}
where $E(\mathbf{\xi})=\mathcal{H}_{\delta}
[\mathbf{S}_{0}(z,\boldsymbol{\xi})]$ is the energy of the domain wall
\cite{Liu11} as a function of the soft modes $\boldsymbol{\xi}$.

For thin nanowires, a DW is a rigid spin texture. Its slow dynamics
can be described in terms of only two collective coordinates
corresponding to zero modes of motion.  These zero modes are the DW
position $z_{0}$ and its conjugate variable -- the tilt angle $\phi$
for the transverse DW. For the vortex DW, $\phi$ can serve as the
magnetization angle defining the transverse position of the vortex in
the wire \footnote{For details see Ref.~\onlinecite{Tretiakov08,
    *Clarke08}}.  Using the definition of $\mathbf{h}_{\alpha }$,
Eq.~\eqref{eq:h}, and the fact that $\dot{\mathbf{S}}_{0} = -
\dot{z}_{0}\partial_{z}\mathbf{S}_{0} +\dot{\phi} \partial_{\phi
}\mathbf{S}_{0}$ we obtain
\begin{equation}\label{eq:h2}
\mathbf{h}_{\alpha }=
\alpha \dot{\phi }\mathbf{S}_{0}\times \partial_{\phi }\mathbf{S}_{0}
-j\partial_{z} \mathbf{S}_{0}
+(\beta j-\alpha \dot{z}_{0})\mathbf{S}_{0}\times \partial_{z} \mathbf{S}_{0} 
-\mathbf{S}_{0}\times \mathbf{H}
\end{equation}
Here we included uniform magnetic field $\mathbf{H}$ in
$\mathbf{h}_{\alpha }$ by adding the term $-\mathbf{S}_{0}\times
\mathbf{H}$ in \eqref{eq:h2}.

Up to the leading order in small dissipation ($\alpha$ and $\beta$)
the equations of motion become
\begin{eqnarray}
\dot{z}_{0}&=& -\frac{1}{2}\frac{\partial E}{\partial \phi} +j,
\label{eq:eq_z0}
\\
\dot{\phi }&=&\frac{1}{2} \frac{\partial E}{\partial z_{0}} 
-\frac{\alpha a_{zz}}{2} \frac{\partial E}{\partial \phi}
+H +(\alpha-\beta)a_{zz}j.
\label{eq:eq}
\end{eqnarray}
Here for simplicity we have taken $\mathbf{H}$ to be along the wire
(in the $z$ direction) and $a_{zz}=\frac{1}{2} \int dz (\partial_z
\mathbf{S}_0)^2$.  These equations are rather universal and can also
be applied to describe the dynamics of vortex domain walls
\cite{Tretiakov08, *Clarke08} in terms of two collective coordinates
associated with the DW degrees of freedom. The 1/2 in the first term
on the right-hand side of Eqs.~(\ref{eq:eq_z0}) and (\ref{eq:eq}) is a
consequence of the Poisson bracket of the conjugated variables $z_0$
and $\phi$.  The most general derivation of these equations is based
on Poisson brackets and energy dissipation and will be presented
elsewhere~\cite{future}.

Equations (\ref{eq:eq_z0}) and (\ref{eq:eq}) do not depend on details
of the microscopic model. The only required input is the energy
of a {\em static} DW as a function of two parameters $z_{0}$ and
$\phi$. This function can be either calculated by means of an
analytical approximate model, micromagnetic simulations, or can be
measured experimentally for a given wire by a method analogous to the
one described in Ref.~\onlinecite{Liu11}.

The energy of a static DW, $E (z_{0},\phi) = \mathcal{H}
[\mathbf{S}_0(z;z_{0},\phi)]$, where $\mathbf{S}_0$ is a solution of a
static LLG, in general depends on both $z_{0}$ and $\phi$. The main
dependence of $E$ on the angle $\phi$ comes from the anisotropy in the
transverse plane, $E (\phi) = -\kappa \cos(2\phi)$~\footnote{For the
  model Hamiltonian of Ref.~\onlinecite{Tretiakov_DMI} this constant
  is $\kappa =\pi K\Gamma \Delta^2/\sinh(\pi \Gamma \Delta)$ which
  reduces to $\kappa =K \Delta$ for $\Gamma\Delta\ll 1$. Here $K$ is
  the transverse anisotropy constant and $\Delta$ is the DW width.}.
The dependence of $E$ on the DW position $z_{0}$ may come from
different sources such as, $z$-dependence of the wire shape,
nonuniform concentration of impurities, wire surface roughness,
nanofabricated notches, etc.  Equations (\ref{eq:eq_z0})
and~(\ref{eq:eq}) can be used to study DW propagation in disordered
wires as well as DW depinning dynamics~\cite{Pi2011, Min10} under the
action of time-dependent magnetic fields and currents.

Taking $H=\partial_{z_{0}}E=0$ in Eq.~\eqref{eq:eq} one
recovers the DW dynamics equations for a translationally invariant
nanowire with no magnetic field applied~\cite{Tretiakov_DMI}. In this
case the DW moves with a constant velocity $\beta j/\alpha$ for small
currents $j$, whereas above the critical current
$j_c=\kappa\alpha/|\alpha-\beta|$ in addition to moving along the wire
its angle $\phi$ rotates around the wire axis.

\textit{Memory device.} As an example of the use of equations
(\ref{eq:eq_z0}) and~\eqref{eq:eq} we consider a magnetic memory
device based on a flat hourglass-shaped nanostrip (shown in
Fig.~\ref{fig:pulse}) with the constant thickness $h$. We propose a
nonvolatile device, which employs the magnetization direction within
the DW as the information storage \footnote{A related idea of dynamics
  of a trapped DW in a spin-valve geometry with current perpendicular
  to the plane has been addressed in
  Ref.~\onlinecite{Mryasov06}.}. Without current, the transverse DW
stays at the place where the nanostrip's cross-section is the
narrowest. When a particular current pulse is applied, the DW
magnetization angle $\phi$ flips from $0$ to $\pi$. At the
intermediate step of this flipping process, the DW also deviates from
the narrowest position of the nanostrip and at the end it comes back
but with the opposite magnetization direction (for computational
details see Fig.~\ref{fig:current_z0_theta} below). The same current
pulse at a later time can move it back to the original
configuration \footnote{See Supplemental Material at http://link.aps.org/supplemental/10.1103/PhysRevLett.108.247201 for a movie demonstrating the mechanism of this memory element}.

The time it takes to switch the magnetization depends on the current pulse
shape. During this process the main energy loss in a realistic wire is the
Ohmic loss. How much energy is needed for a single switching also
depends on the parameters of the current pulse.

\begin{figure}
\includegraphics[width=0.9\columnwidth]{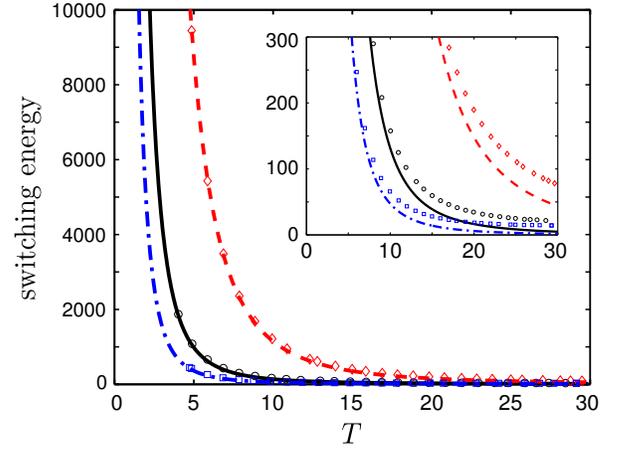}
\caption{(color online) Dependence of the switching energy in units of
  $\int^T_0 j^2(t)\, dt$ on the switching time $T$ for $\sigma = 0.01$
  (red dashed line), $\sigma = 0.03$ (black solid line) and $\sigma =
  0.05$ (blue dot-dashed line). The inset shows the same dependence
  for smaller range of the switching energies. One can clearly see the
  deviations of the simulated data from the analytical solution for
  the small $T$ limit.}
\label{fig:switching_energy}
\end{figure}

We, thus, aim to solve the following problems: 1) What is the
optimal (requiring the least amount of energy) current pulse shape for
a given switching time? 2) How the minimal required energy per flip depends
on the switching time? The answers to these questions are given in
Figs.~\ref{fig:switching_energy} and~\ref{fig:current_z0_theta}. Here
we sketch the calculation.

For a smooth nanostrip, we can approximate its width by a parabolic
shape $w(z )=w_0 + z^2/R$ and its thickness $h\ll w_0$, see
Fig.~\ref{fig:pulse}. Then the correction to the DW energy due to the
shift from the center of the strip becomes $E_0 (z_0)= \frac{\gamma_0
  J}{M w_0 \Delta R}z_0^2$, where $J$ is the exchange constant,
$\Delta$ is the DW width, and $R$ is the curvature radius of the
nanostrip.

Taking $H=0$ (no magnetic field applied) and
$a_{zz}=1/\Delta$ \footnote{This is true for the systems without
  Dzyaloshinskii-Moriya interaction (DMI). If DMI is present, $a_{zz}$
  acquires correction: $a_{zz}=(1+\Gamma^2\Delta^2)/\Delta$ where
  $\Gamma=D/J$ and $D$ is the DMI constant, see
  Ref.~\onlinecite{Tretiakov_DMI}.}, and rescaling variables as
$t\rightarrow \kappa t/\Delta$, $z_0 \rightarrow z_0/\Delta$, and
$j\rightarrow j/\kappa$ to make all of them dimensionless,
Eqs.~(\ref{eq:eq_z0}) and (\ref{eq:eq}) become
\begin{eqnarray}
&&j=\dot{z}_{0}+\sin 2\phi,
\label{eq:j} \\
&&\Omega (t)\equiv \dot{\phi} -(\alpha -\beta )\dot{z}_{0}
+\beta \sin 2\phi -\sigma z_{0}=0.
\label{eq:constraint}
\end{eqnarray}
Here we have used the fact that in dimensionless variables the
$z_0$-dependence of the DW energy is $E_0 (z_0)=\sigma z_0^2$ with
\begin{equation}
\sigma = \frac{\gamma_0 J \Delta }{M w_0 \kappa R}.
\end{equation}
We can estimate $\sigma$ for materials such as Permalloy using the
gyromagnetic ratio $\gamma_0 = 1.76\times 10^{11}$ s$^{-1}$T$^{-1}$,
the exchange constant $J = 1.3\times 10^{11} $ J/m, the saturation
magnetization $M = 8 \times 10^5$ A/m, and $\kappa = j_c
(\alpha-\beta)/\alpha$ where $(\alpha-\beta)/\alpha \approx 1$ and the
critical current measured in units of velocity $j_c \approx 100$
m/s. It gives $\sigma \approx \Delta/(w_0 R)\times 10$ nm. Further
taking for realistic nanowires $R = 100$ nm, $w_0= 100$ nm, and
$\Delta \approx 10$ nm, we find that $\sigma \approx 0.01$ which makes
it of the order of $\alpha $ and $\beta$. It also justifies neglecting
the terms proportional to $\alpha\sigma$ and $\beta\sigma$ in
Eqs.~(\ref{eq:j}) and (\ref{eq:constraint}).

Our goal is to minimize the Ohmic losses~\cite{Tretiakov:losses,
  Tretiakov:JAP} per one switch of the memory bit. It corresponds to
the flipping of the DW angle $\phi$ between two stable values $0$ to
$\pi$, which are defined by the minima of transverse anisotropy. To
achieve this, one has to minimize $j^{2}$ during the switching time
$T$ while keeping the constraint \eqref{eq:constraint}.

The fact that the DW is at rest at the angle $\phi=0$ before the
current pulse and at rest again at the angle $\phi=\pi$ immediately
after the current pulse is taken into account by the boundary
conditions $\phi (0)=0$, $z_{0}(0)=0$, $\dot{z}_{0}(+0)=0$, and
$\phi (T)=\pi$, $z_{0}(T)=0$, $\dot{z}_{0}(T-0)=0$.

In order to find minimum of the power $\int_{0}^{T}j^{2}dt$ with
constraint \eqref{eq:constraint}, which must hold at all times, we use
the Lagrange multiplier $\dot{J}(t)$ and minimize the functional
\begin{equation*}
\label{eq:functional}
\int_0^T [j^2 - \dot{J}(t)\Omega(t)] dt = \int_{0}^{T}\!\left[\left( \dot{z}_{0}+\sin 2\phi \right)^{2}
-\dot{J}(t)\Omega (t)
\right] dt
\end{equation*}
with respect to three functions $z_{0}(t)$, $\phi (t)$, and
$\dot{J}(t)$ with $J(0)=0$. Then, in addition to
Eq.~(\ref{eq:constraint}), we obtain two equations
\begin{eqnarray}
&&\ddot{J }
-2\beta \dot{J} \cos 2\phi + 4(\dot{z}_{0}+\sin 2\phi )\cos 2\phi =0,
\label{eq:eq2J}\\
&&j = j_{0}+\frac{\sigma }{2}J-\frac{\alpha -\beta }{2}\dot{J },
\label{eq:jJ}
\end{eqnarray}
where to write the last equation we used Eq.~(\ref{eq:j}), and $j_{0}$
is an integration constant. It can be seen that $j_{0}=j(0)$ if
$\dot{J}(0)=0$.

First, we consider the case when the current is absent,
$j=\dot{z}_{0}+\sin \theta =0$. Then $J=0$ and equation \eqref{eq:constraint}
gives
\begin{equation}
\label{eq:oscillations}
\ddot{\phi} +2\alpha \dot{\phi }\cos 2\phi +\sigma\sin 2\phi =0.
\end{equation}
For small angles $\phi$, this equation describes a damped harmonic
oscillator. For $\alpha \ll \sqrt{\sigma}$ the motion is underdamped
and the DW performs oscillations with the frequency $\omega_{0}
=\sqrt{2\sigma}$. For $\alpha \gg \sqrt{\sigma}$ the motion is
overdamped.

If one kicks the DW with a very narrow current pulse $j(t)=A\delta
(t)$, integrating $\dot{z}_{0}+\sin 2\phi =A\delta (t)$ from
$-\epsilon $ to $\epsilon $ with the initial condition
$z_{0}(-\epsilon )=0$ and taking the limit $\epsilon \rightarrow 0$,
we find $z_{+}=\lim\limits_{\epsilon \rightarrow 0}z_{0}(\epsilon
)=A$. The same procedure for equation \eqref{eq:constraint} gives
$\phi_{+}=(\alpha -\beta )z_{+}=(\alpha -\beta )A$.  After this pulse,
the motion will be described by equation \eqref{eq:oscillations}.
Depending on the characteristic time scales in this equation,
$1/\sqrt{\sigma }$ or $1/\alpha$, the motion will be underdamped or
overdamped.

Let us consider the limiting case $\alpha =\beta =0$ of the system
without dissipation. Using Eqs.~\eqref{eq:j}, \eqref{eq:constraint},
and \eqref{eq:jJ} we find
\begin{eqnarray}
&&z_{0}=\frac{\dot{\phi}}{\sigma },
\label{eq:eq3Ja0b0}\\
&&J = \frac{2}{\sigma }\left( \frac{\ddot{\phi }}{\sigma}
+\sin 2\phi -j_{0}\right).
\end{eqnarray}
Substituting these equations into Eq.~\eqref{eq:eq2J} we also obtain
\begin{equation}
\label{eq:One}
\left( 2\sigma \cos 2\phi +  \frac{\partial^{2}}{\partial t^{2}} \right)
\left(\sigma \sin 2\phi +  \frac{\partial^{2}}{\partial t^{2}} \phi \right)
=0.
\end{equation}
There is a trivial solution of this equation, $\phi (t) = 0$, but it
does not satisfy the boundary conditions: the angle does not flip in
this case.  However, in the limit when the switching process is fast,
or alternatively $\sigma$ is small, one can find an approximate
solution.  If the switching time $T$ satisfies $T\sqrt{\sigma }\ll 1$,
all the terms except for the fourth derivative can be neglected and we
obtain $\partial^{4} \phi/\partial t^{4} =0$.  Using the boundary conditions for
both $\phi$ and $z_{0}$ one then finds
\begin{equation}
\label{eq:theta}
\phi (t)=3\pi (t/T)^{2}-2\pi (t/T)^{3}
\end{equation}
which gives
\begin{equation}
\label{eq:jTheta}
j(t)=\frac{1}{\sigma }\ddot{\phi}
+\sin 2\phi\approx \frac{1}{\sigma }\ddot{\phi}
=\frac{6\pi }{\sigma }\left(1-\frac{2t}{T}\right).
\end{equation}

The switching energy measured in dimensionless units $\int^T_0
j^2(t)\, dt$ is shown in Fig.~\ref{fig:switching_energy}. For small
$T$, according to Eq.~(\ref{eq:jTheta}), it is given by
$12\pi^2/(\sigma^2 T^3)$. The actual units of the switching energy can
be estimated as $\rho J_c^2 \Delta/(j_c h)$, where $\rho\approx
1.4\times 10^{-7}\, \Omega \rm{m}$ is Permalloy resistivity, the
critical current density $J_c \approx 10^{12}$ A/m$^2$, $\Delta\approx
h \approx 10$ nm; and the units of time are $\Delta/j_c \approx
10^{-10}$ s. This gives the units of the switching energy to be
$\approx 10^{-16}$ J.  For comparison, best MRAM devices typically
consume $10^{-10}$ J per switching a bit~\cite{Everspin_web} in 5
ns. This shows that our proposed memory device is about $10^{5}$ times
more energy efficient for the same switching
times. Figure~\ref{fig:switching_energy} also gives the estimate of
the best switching time achievable for any particular energy supplied
per switch.

\begin{figure}
\includegraphics[width=1\columnwidth]{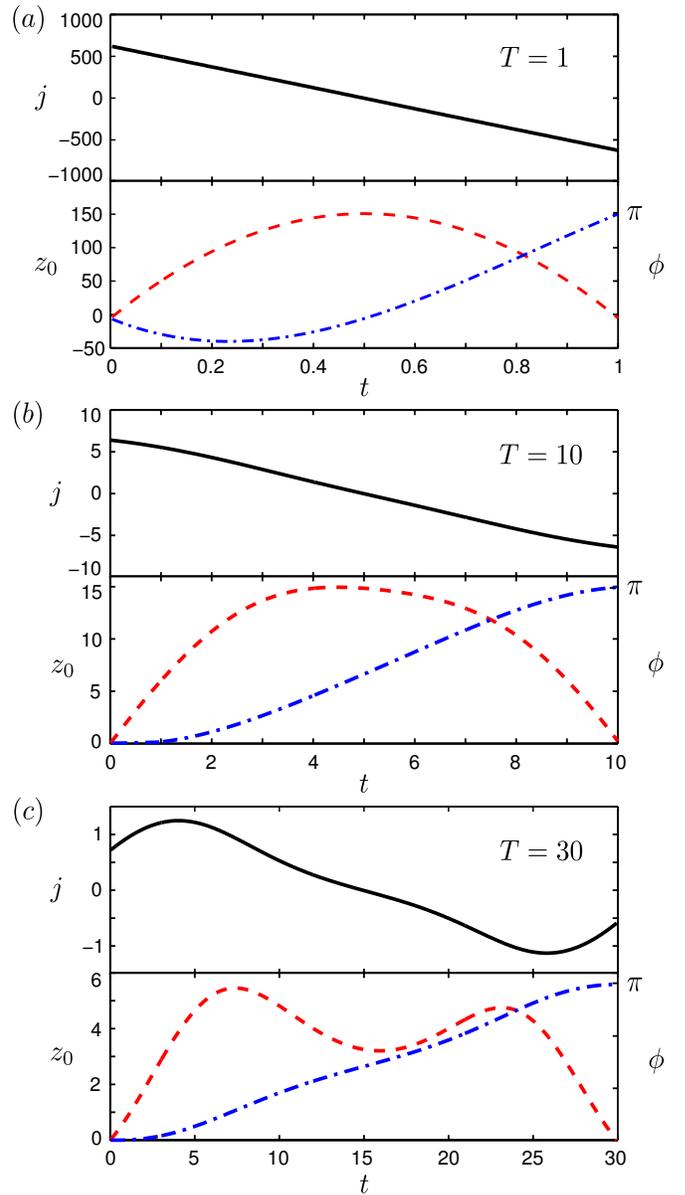}
\caption{(color online) The time dependence of the DW
  position and angle for the optimal current pulse with the
  parameters $\alpha = 0.01$, $\beta = 0.02$, and $\sigma =
  0.03$. Optimal current $j(t)$ (black solid line), $z_0(t)$ (red
  dashed line) and angle $\phi(t)$ (blue dash-dotted line) as
  functions of time $t$ for (a) $T=1$, (b) $T=10$, (c) $T=30$. }
\label{fig:current_z0_theta}
\end{figure}

The full set of equations with realistic $\alpha $ and $\beta $
parameters can be solved numerically.  The comparison of the
analytical solution~(\ref{eq:jTheta}) for small $T$ with the numerics
for a range of $T$ is shown in Fig.~\ref{fig:current_z0_theta}. We
find that the linear current coincides with the simulation result for
the short switching times, i.e. $T\lesssim 1/\sqrt{\sigma}$. For long
switching times, the optimal current $j(t)$ flips the DW and
then lets it relax with some oscillations to $z_0 = 0$ and $\phi =
\pi$ within the required time $T$.

\textit{Summary}. We have generalized the DW dynamics to the case of
translationally non-invariant ferromagnetic nanowires. The obtained
equations of motion make the description of the DW propagation closer
to experiment by accounting for smooth surface roughness and other
disorder effects. We have also considered an hourglass-shaped
nanostrip with a transverse DW trapped in the middle as a prototype of
a magnetic memory device.  The exact time-dependence of the current
pulses required to switch the magnetization with the minimal Ohmic
losses per switching has been obtained. Furthermore, we find how the
switching time depends on the Ohmic losses per switching for the
optimal current pulse.  Our estimates show that this hourglass-shaped
nanodevice may be $10^5$ times more energy efficient for the same
switching times as used in the best modern memory devices.

This work was supported by the NSF Grant No. 0757992,
ONR-N000141110780, and Welch Foundation (A-1678).

\bibliography{magnetizationDynamics}

\end{document}